# The Architecture of Inter-Level Representation

**Harry Sticker**
**Ganymede Technology**
New York, New York, USA
hsticker@ganymedetechnology.com

## Abstract

Inter-level connections in science routinely require constructs that neither of the connected theories contains. Statistical mechanics needs assumptions such as the Stosszahlansatz to generate thermodynamic irreversibility—assumptions Hamiltonian mechanics cannot supply. Quantum chemistry offers four incompatible analyses of chemical bonding for the same quantum state, none of which are selected by the Schrödinger dynamics. Molecular genetics has not converged on a stable definition of the gene despite decades of molecular detail. These are not isolated anomalies but instances of a common architectural pattern.

The missing apparatus is the bridge theory: a third theoretical role that connects a dynamical theory to an observational theory through a many-to-one inter-level map. That map generates the contingent space—the set of dynamical states compatible with an observational description but not selected by it—whose geometry neither connected theory determines. Completing the bridge theory requires three conditions in order: a Partition that defines observational equivalence classes; a Magnitude that characterizes the geometry and scale of the contingent space; and a Closure that selects or weights its elements.

The framework yields an objective distinction between closing and introducing rules, formalized by the Mirror Test, and supports a tripartite taxonomy of emergence. It explains why some inter-level disputes persist and what would be required to resolve them.

# 1. The Third Role

## 1.1 The Recurring Pattern

In 1876, Loschmidt showed Boltzmann that his H-theorem proved too much. For every molecular trajectory along which entropy increases, the time-reversed trajectory is equally admissible by the Hamiltonian and shows entropy decreasing. The derivation was correct. The conclusion was impossible. Something in the connection between Hamiltonian mechanics and thermodynamics was doing work that neither theory acknowledged.

The same structural gap appears wherever inter-level connections have been seriously pursued. In quantum chemistry, the Schrödinger equation determines the electron density but does not say how that density constitutes a chemical bond: four incompatible analysis schemes answer the question differently, and quantum mechanics does not adjudicate among them. In molecular genetics, 70 years of molecular knowledge have not produced a stable definition of the gene because molecular biology does not settle which features of DNA count as "the gene" for a given explanatory purpose. In each case, the connecting apparatus requires constructs that neither connected theory contains nor can supply. What those constructs are, why they cannot be derived, and what structure they must exhibit are the questions this paper addresses.

**Statistical mechanics.** Connecting Hamiltonian mechanics to thermodynamics, statistical mechanics required Boltzmann to model the thermodynamic level by partitioning phase space into macrostates—a coarse-graining that the molecular Hamiltonian does not determine. From this, he derived the H-theorem for the monotonic growth of a measure he considered a counterpart of entropy. His derivation, however, requires the Stosszahlansatz: the assumption that molecular velocities before collision are statistically independent. This assumption introduces a temporal preference absent in Hamiltonian dynamics—one that, as Loschmidt showed in 1876, cannot be derived from time-symmetric mechanics. For any trajectory along which the contingent space expands, the time-reversed trajectory is dynamically admissible and shows contraction. Subsequent approaches—the ergodic program and typicality—aimed to eliminate these constructs without success (Boltzmann 1877; Loschmidt 1876; Earman and Rédei 1996; Sklar 1993; Dizadji-Bahmani et al. 2010).

**Quantum chemistry.** The development of quantum mechanics in the 1920s raised the prospect of deriving the properties of chemical bonds directly from the Schrödinger equation—a prospect that has not been fulfilled, not because quantum mechanics is incomplete, but because the connection to bonding is. Molecular orbital theory, valence bond theory, natural bond orbital analysis, and atoms-in-molecules analysis each provide different, sometimes incompatible descriptions of bonding for the same quantum state. This incompatibility is not due to computational approximation but to a foundational disagreement about which quantum-mechanical quantities constitute chemical bonds—one that persists after the quantum state has been fully specified. The Schrödinger dynamics do not select among the analysis schemes: the underdetermination is structural, not a symptom of incomplete quantum mechanics (Chang 2012; Bader 1990; Hoffmann et al. 2003; Hendry 2010).

**Molecular genetics.** Gregor Mendel's seminal work in 1866 established the existence of discrete hereditary factors without any knowledge of their physical substrate. When molecular biology provided that substrate a century later, it raised the prospect of a stable, molecularly grounded definition of the gene. That prospect has not been fulfilled. A straightforward identification of genes by DNA sequencing is complicated by open reading frames, regulatory sequences, overlapping genes, and RNA genes, which—from the cistron to the transcription unit to the genomic region with a heritable phenotypic effect—frustrate convergence. Different explanatory demands yield incompatible characterizations of what a gene is, none of which is privileged by the underlying molecular biology (Waters 1994; Griffiths and Neumann-Held 1999; Kitcher 1984).

In each case, the same question presses: is the failure to derive these constructs a sign of theoretical immaturity, or does it reflect something principled about the architecture of inter-level connections itself? The architecture of the bridge theory answers it.

## 1.2 Why the Two-Theory Picture Cannot Answer It

Nagel (1961) identified the central problem: reduction requires bridge laws connecting the vocabularies of the two theories. But where does a bridge law belong? Assign it to the upper theory, and it inherits that theory's explanatory obligations, which it cannot discharge without the lower theory. Assign it to the lower theory, and the same difficulty recurs in reverse. Leave it unassigned, and the reduction has an unexplained middle term. Every subsequent repair—Schaffner (1993), Dizadji-Bahmani, Frigg, and Hartmann (2010)—kept the auxiliary constructs inside one of the two connected theories, where they acquired that theory's dynamical obligations and could not discharge them. The two-theory picture has no structural slot for constructs that belong to neither connected theory.

The emergence literature pressed on the same gap from the other direction. Call something weakly emergent if it is "derivable in principle, not yet derived." Call something strongly emergent if it is "not derivable, for unspecified structural reasons." A reductionist and an anti-reductionist can now classify thermodynamic irreversibility differently—and neither can be shown to be wrong. The binary generates disagreements without resolving them because it supplies no criterion: it names the distinction it cannot draw. A vocabulary without a criterion is not a classification. It is a filing system. Kim (1998) recognizes that supervenience-based accounts presuppose a determinate relationship between levels, but nothing in the supervenience framework specifies what that relationship must contain. Batterman (2002) makes a genuine advance, locating emergence in the asymptotic structure of limiting relations between theories—but the apparatus stays internal to the dynamical theory. It identifies where emergence happens. It does not identify the constructs that do the work.

Butterfield (2011) argues that emergence and reduction are compatible when emergent behavior arises in a limit: a property can be novel at the macro-level and yet continuously connected to the underlying dynamics. Correct as far as it goes—but the limiting relation is silent on the structural source of novelty. The Mirror Test addresses a distinct question: whether the closure rule required to generate the observed behavior is G-invariant. A property can be emergent in Butterfield's sense and still be classified as provisional or permanent depending on that criterion.

Pearl (2000) comes closest. His do-calculus shows that causal and statistical descriptions play distinct and irreducible roles—neither can be collapsed into the other—but stops short of specifying what the connection between them must entail. That specification is the bridge theory.

**1.3 The Contingent Space**

Every account surveyed above runs into the same omission from a different direction. None provides an apparatus to specify what the connecting constructs contain, how large the space they operate on is, or how it is closed. The omission has a name: the contingent space. When the inter-level map is many-to-one, it generates a contingent space: the set of dynamical states compatible with an observational description but not selected by it. It is not ignorance, nor an artifact of approximation, nor eliminable by redescribing the observations. It is the structural consequence of describing the world at a level coarser than the one at which it is dynamically constituted.

The contingent space is what the two-theory picture provides no slot for—Nagel's bridge laws operate over it implicitly without naming it. Other accounts of reduction or emergence—semantic (van Fraassen 1980; Suppe 1977), mechanistic (Machamer et al. 2000; Craver 2007), or interventionist (Woodward 2003; Strevens 2008)—only hint at it. None provides the apparatus to specify what the contingent space contains, how large it is, or how it is closed.

Theoretical connections between levels involve three distinct theoretical roles: dynamical, observational, and bridge. Positions in inter-level architecture rather than intrinsic properties of theories, these roles can be occupied by the same theory in different connections: thermodynamics plays an observational role relative to statistical mechanics and a dynamical role relative to fluid dynamics. Throughout this paper, "bridge theory" refers to this third structural role—the position in inter-level architecture defined by the inter-level map, the characterization of the contingent space it generates, and the closure rules that complete it.

In the dynamical role, a theory specifies a state space and a flow over it. For Hamiltonian systems, the flow singles out the Liouville measure as a natural invariant measure; for Schrödinger evolution, the unitarily invariant measure. In the observational role, a theory specifies constraint relations between primitives without supplying a generating mechanism. It characterizes admissible configurations without explaining how they arise. In the bridge role, a theory specifies the inter-level map between the dynamical state space and the space of observational descriptions—a specification that generates the contingent space and requires three completeness conditions to characterize it. The bridge theory contains no evolution equations of its own: it filters and constrains dynamical states but does not generate trajectories.

The present account, therefore, differs from familiar discussions of modeling or idealization. A bridge theory is not merely a modeling choice but a structural role in the architecture of inter-level representation, defined by the inter-level map, the contingent space it generates, and the completeness conditions required to characterize and close that space. A modeling choice may or may not generate a contingent space; it has no obligation to satisfy the completeness conditions in order, and it carries no architectural constraints on its closure rules. The bridge theory has all three. The difference is not terminological. It determines what questions a framework can and cannot answer.

## 2. The Structure of the Bridge Theory

### 2.0 The Intuitive Picture

The dynamical theory is Hamiltonian mechanics—it specifies a flow on a phase space of molecular configurations. The observational theory is thermodynamics—it specifies relations among temperature, pressure, volume, and entropy. The bridge theory is the structural role that connects them. Three conditions constitute it.

The first is to decide what counts as the same thing at the observational level. Two molecular configurations that produce the same temperature are "the same" for thermodynamic purposes—but that sameness is not given by the physics; it is chosen. The Partition—the first constitutive condition of the bridge theory—fixes which macroscopic quantities matter, at what scale, and for what explanatory purposes, and thereby generates the contingent space: the set of all molecular configurations compatible with a given macrostate. Different Partition choices yield distinct contingent spaces and distinct bridge theories. A bridge theory that has not fixed its Partition has not yet defined its subject matter.

The second is to characterize the geometry of the contingent space—not just which configurations belong to it, but their relative weight and structure. Knowing the members of the contingent space is not enough. A gas at high temperature has a vastly larger contingent space than one at low temperature. Without knowing the geometry, one cannot say which configurations are typical, why entropy increases rather than decreases, or why Mendel's 3:1 ratio follows necessarily from his rules rather than being a coincidence. The contingent space provides membership; Magnitude provides structure. Knowing that a room contains furniture tells one nothing about whether it is crowded.

The third is to specify how the contingent space closes onto the observations. Even with a fully characterized contingent space, nothing selects which elements are realized or how weight is distributed over them. But there is a deeper problem. The dynamical theory is symmetric—time-reversal symmetric in the Hamiltonian case, unitarily symmetric in the quantum case. The observations it is meant to explain often are not: thermodynamic irreversibility picks out a direction in time, and chemical bonds pick out specific spatial locations. When the observations require a symmetry-breaking selection, the closure rule must supply it—something the dynamical theory, precisely because it is symmetric, cannot do. That is the problem the Mirror Test is designed to detect.

The three conditions are not arbitrary. Each follows from the structure of the contingent space itself: one cannot characterize its geometry without first defining it, and one cannot close it without having characterized it. The ordering is a constraint imposed by the architecture, not a choice. The contingent space, the object all three conditions are about, is what the two-theory picture has no slot for.

## 2.1 The Inter-Level Map

When a theory describes a system at a coarser level than the one at which its dynamics operate, the connection is an inter-level map: a function $\pi$ from the dynamical state space to the space of observational descriptions. Two structural features of this map shape everything that follows.

$\pi$ is always many-to-one: the observational description discards information the dynamical level carries, so multiple distinct dynamical states map to the same description. This is what Fodor (1974) identified as multiple realizability—the same higher-level description is compatible with many distinct lower-level realizations. In statistical mechanics, every molecular configuration maps to a single macrostate characterized by temperature, pressure, and volume (Boltzmann 1877; Tolman 1938). In quantum chemistry, many quantum states are compatible with a given chemical structural description, and which description applies depends on the analysis scheme used to partition the electron density (Bader 1990; Hoffmann et al. 2003).

The many-to-one structure of $\pi$ has one further consequence worth making explicit. The mapping from dynamical states to observational descriptions is compressive: it discards information that the dynamical level carries but the observational level does not. That compression is directional and asymmetric: going from dynamical states to observational descriptions loses information, while recovering dynamical states from an observational description requires the full contingent space. This structural asymmetry is a necessary precondition for entropy increase to be representable at the bridge level. A many-to-one map generates a contingent space in which the volume of dynamical states compatible with a given observational description can grow under the dynamical flow; no such growth is possible if $\pi$ is injective. The Closure rules discussed in §2.4 and the Mirror Test introduced in §3 operate against this background: they determine which direction the compressive map runs and which temporal direction entropy increase tracks, but they presuppose the asymmetry that the many-to-one structure of $\pi$ already supplies.

$\pi$ also always instantiates a Partition—the bridge-theoretic condition that specifies which macroscopic quantities are tracked, at which observational scale, and for what explanatory purposes. The coarse-graining is the most fundamental bridge-theoretic construct; everything else in the bridge theory presupposes it. The coarse-graining of gas distributions is relatively uncontroversial; the coarse-graining of chemical bonding is actively contested: different analysis schemes make different choices about how to partition the electron density, generating different contingent spaces for the same quantum state. A bridge theory that has not specified the form of coarse-graining has not yet defined its subject matter.

## 2.2 The Contingent Space

The inter-level map is many-to-one, so every observational description has a pre-image—the contingent space: the set of dynamical states that produce it. Formally, for any inter-level map $\pi$, $C(x) = \pi^{-1}(x)$ is the set of dynamical states mapping to the observational description x. For a thermodynamic system in macrostate $M = (T, P, V)$, this is the region $\Gamma_M$ of phase space—i.e., all molecular configurations that produce the same thermodynamic description while differing in every individual molecular detail.

The contingent space cannot be derived from either connected theory. Given only the observational description, the dynamical theory cannot select among the elements of $C(x)$: any element is dynamically admissible. It cannot be eliminated by redescription either: any two states in $C(x)$ are observationally equivalent, so no refinement of the observations can distinguish them without changing the Partition itself. And it cannot represent a single realized history: the observational description is consistent with multiple dynamical histories and does not select one. These are not limitations of our knowledge. They are structural consequences of the many-to-one architecture of the inter-level map.

A fourth observation concerns the metaphysical standing of the contingent space. The framework is neutral on whether $C(x)$ represents genuine physical indeterminacy—a real plurality of dynamical states, each of which could be actual—or epistemic limitation about a determinate underlying state that has been coarse-grained away. Different theoretical traditions read it differently. In classical statistical mechanics, the standard reading is epistemological: one microstate is actual, and $C(x)$ represents ignorance of which. In quantum mechanics, the question is contested, and different interpretations disagree about whether the contingent space before closure is ontologically or merely epistemologically indeterminate. The bridge theory framework does not resolve this disagreement. Its task is to characterize the geometry of $C(x)$—its size, structure, and closure conditions—not to determine what kind of fact constitutes it. This neutrality is not a deficiency: the framework can accommodate theories that take opposing positions on the metaphysics of contingent space, while the Partition, Magnitude, and Closure conditions remain in force regardless of the position taken.

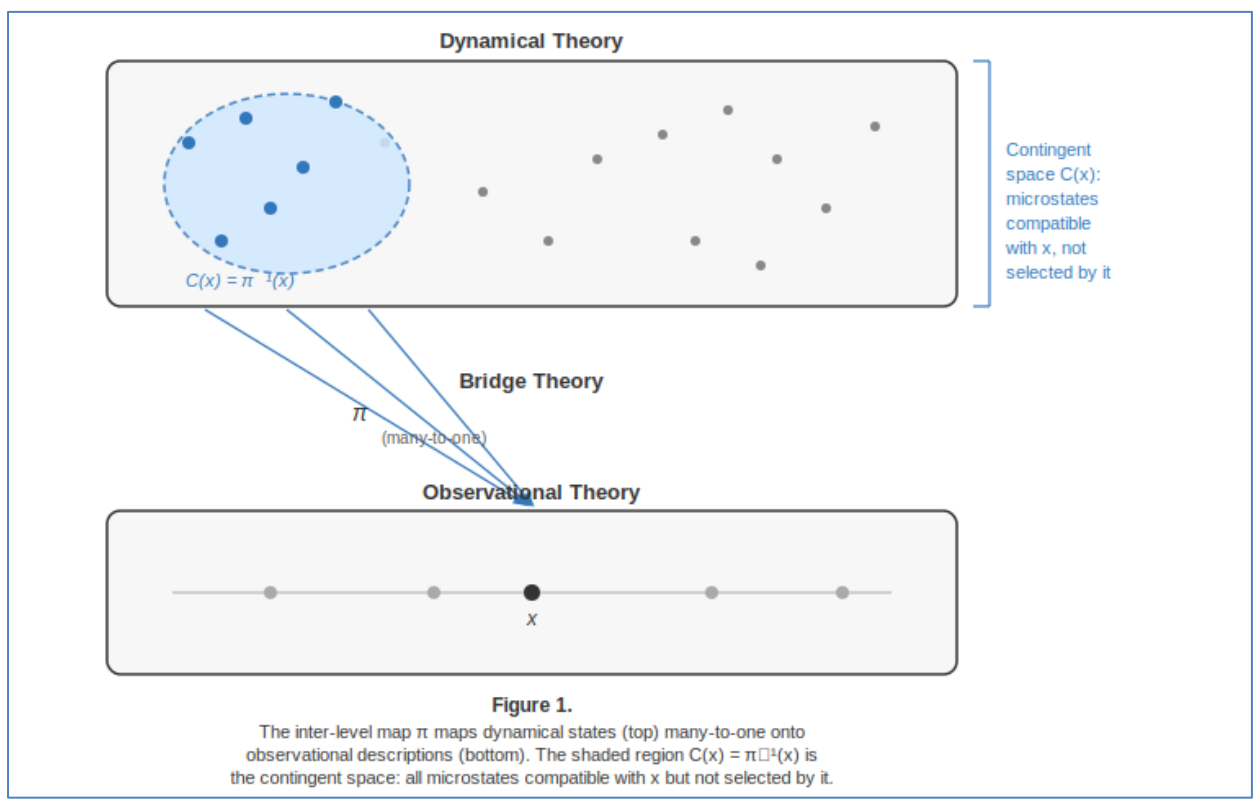

**Figure 1.**
The inter-level map π maps dynamical states (top) many-to-one onto observational descriptions (bottom). The shaded region C(x) = π⁻¹(x) is the contingent space: all microstates compatible with x but not selected by it.

A pair of dice makes the structure concrete. Consider a single die. The dynamical theory describes the physical trajectory of the toss—the initial velocity, angular momentum, and surface interactions that determine where it lands. The observational description is coarser: it records only the face showing. The Partition maps every physical trajectory to one of six outcomes. The contingent space for "a three" is the set of all physical trajectories that land on three—a large and heterogeneous set, none of whose members is selected by the observational description.

Now consider two dice and the sum showing. The observational description "seven" is compatible with six physical combinations: (1,6), (2,5), (3,4), (4,3), (5,2), (6,1). The observational description "two" is compatible with only one: (1,1). The contingent spaces have different magnitudes. That difference is an intrinsic structural feature of the contingent space—it exists before any probability has been assigned. The reason seven is more probable than two is not assumed; it follows from the difference in Magnitude once Closure distributes weight uniformly over the physical combinations. Probability is what Closure introduces. The contingent space, and its Magnitude, come first.

## 2.3 The Contingency Measure

The dice example establishes that the contingent spaces for "seven" and "two" have different sizes, and that difference does the explanatory work. Measuring that size precisely turns out to require two components of different character—one supplied by the dynamical theory, one that the dynamical theory cannot supply.

For an observational description x, let $W(x)$ denote the volume of its contingent space in units of a reference cell $v_0$. This generalizes Boltzmann's W from a count of discrete microstates to a continuous phase space volume with an explicit reference cell. The contingency measure is the logarithm of $W$:

$$cm(x) = \log W(x) = \log(|C(x)| / v_0)$$

The logarithm is taken for additivity, following Boltzmann, Shannon, and others. For two independent systems, contingent spaces combine multiplicatively: $W(x_1, x_2) = W(x_1) \cdot W(x_2)$. Contingency measures, by contrast, add: $cm(x_1, x_2) = cm(x_1) + cm(x_2)$.

More precisely, the contingency measure is a functional of the inter-level map:

$$cm(x) = \log[(1/v_0) \int \delta(\pi(\gamma) - x) \, d\gamma]$$

where the delta function enforces the coarse-graining constraint—selecting precisely those dynamical states $\gamma$ that project to observational description x under $\pi$—and $d\gamma$ is the Liouville measure supplied by the dynamical theory. The formula makes the dependence on both bridge-theoretic constructs visible in the structure of the integral itself: change the inter-level map $\pi$ and the domain of integration; change the measure $d\gamma$ and the geometry of that domain. These are distinct contributions of distinct character.

The formula exposes two components of different character. The first component is the Liouville measure $d\gamma$, the unique invariant volume form singled out by the Hamiltonian flow. It is not a bridge-theoretic choice: the dynamics of the dynamical theory single it out independently of any bridge-level commitment. The second component is the reference cell $v_0$, required to make the argument of the logarithm dimensionless. Hamiltonian mechanics determines the geometry of the contingent space but provides no natural scale to fix the size of an elementary cell. The reference cell is therefore a genuine bridge-theoretic construct, distinct from the Liouville measure and not derivable from the dynamical theory alone. When the dynamical theory is embedded in a more fundamental framework that provides a natural scale, that scale determines $v_0$, and the contingency measure acquires physical content. When it is not further embedded, $v_0$ must be stated explicitly, or contingency measure differences must be used in place of absolute values.

Three structural consequences follow from the definition. First, differences in contingency measures are independent of $v_0$ and always well-defined: the reference cell cancels in any difference, so all comparative claims hold regardless of $v_0$. Second, absolute contingency measures are $v_0$-dependent, unless a framework outside the classical bridge theory resolves $v_0$. Third, the Liouville measure $d\gamma$ is not probabilistic: it characterizes the geometry of the contingent space, and interpretations in terms of probability occur only when a closure rule is provided that assigns weights over the geometry of the space (Gibbs 1902).

A further structural consequence concerns typicality approaches in statistical mechanics. Typicality arguments proceed by showing that a given behavior holds for almost all microstates compatible with a macrostate, measured by the Liouville measure. Such arguments are formulated as ratios of contingent space volumes: the fraction of $\Gamma_M$ for which a given behavior obtains. In any such ratio, $v_0$ cancels: the reference cell appears in both numerator and denominator and drops out. Typicality therefore does not resolve $v_0$—it operates in a regime where $v_0$ is irrelevant to the result. This is not a limitation of typicality but a precise characterization of its scope: typicality conclusions are $v_0$-insensitive by construction, and they hold regardless of how $v_0$ is fixed. The typicality program occupies a position at Closure—establishing what holds for the overwhelming majority of the contingent space—not at Magnitude, where the scale of that space is determined.

## 2.4 The Three Completeness Conditions

The three completeness conditions must be satisfied in order because each presupposes the one before it. Specify a closure rule before the contingent space has been characterized, and you are setting conditions on an object whose geometry is unknown. Fix the geometry before the Partition has been defined, and you are measuring a space whose membership is indeterminate. The ordering is a constraint imposed by the architecture, not a convention: a bridge theory that has not satisfied all three has not finished its job.

The first condition, Partition, specifies what counts as observationally equivalent: which macroscopic quantities individuate the description, at what scale, and for what explanatory purposes. It defines the equivalence classes of the inter-level map and thereby constitutes the contingent space. Without it, the bridge theory has no subject matter. Different Partition choices for the same dynamical theory generate distinct contingent spaces, distinct values of the contingency measure, and distinct closure requirements.

The second condition, Magnitude, characterizes the size and scale of the contingent space. It has two components of different character. The Liouville measure $d\gamma$ is singled out by the dynamics of the dynamical theory and is not a bridge-theoretic choice. The reference cell $v_0$ must be supplied as a genuine bridge-theoretic construct. As defined in §2.3, the contingency measure $cm(x) = \log(|C(x)| / v_0)$ captures both components: $d\gamma$ supplies the geometry of the contingent space, and $v_0$ supplies the scale. Without both components, the bridge theory either lacks the geometry of the contingent space or cannot express that geometry as a dimensionless quantity. Closure rules proposed before both components are specified impose conditions on a space whose magnitude has not been fully established.

The third condition, Closure, specifies how the contingent space is closed. It takes one of two forms: state restriction, which selects a subset of the contingent space as the admissible realizations; or weight assignment, which distributes a measure over the full contingent space. The Stosszahlansatz is a closure rule of the first kind; the microcanonical ensemble is a closure rule of the second. Probability is one species of weight assignment.

## 2.5 The Necessity of Each Condition

Mendel's (1866) paper is the clearest demonstration that Magnitude is a genuine independent condition. Mendel had a precise Partition, in the inferred factors mapping to phenotypic traits, and formally correct Closure rules that gave him the 3:1 and 9:3:3:1 ratios he observed. What was missing was the Magnitude of the contingent space. The 3:1 statistical ratio holds only if the inferred factors are discrete physical units that are segregated with equal weight. For Mendel, these were abstract inferences—he did not know how many such factors exist, whether they are genuinely discrete, or whether the factors are, in fact, the correct ones. Neither the geometry of the factor space nor the natural reference cell had been physically established. This gap prevented completion for 34 years. The cytological work of the 1880s and 1890s supplied both simultaneously: chromosome structure established that factors are discrete physical objects that come in pairs, and the discreteness of chromosomal units provided the natural reference cell against which combinatorial ratios could be measured. The chromosome is the $v_0$ of Mendelian genetics: it is the elementary unit that makes the contingent space countable rather than merely characterizable, and against which the combinatorial ratios acquire their determinate values. Magnitude is not a placeholder between Partition and Closure. It is the condition under which Closure rules have a geometry and a scale to operate on.

Without a Partition, the contingent space is undefined. Caloric illustrates the failure mode: the bridge theory collapsed at Partition because the equivalence classes it required—states at the same "caloric content"—were not stable under Hamiltonian dynamics. When the Partition generates equivalence classes that the dynamics immediately scramble, the bridge theory has not defined its subject matter. It has no subject matter to name, no Magnitude to characterize, no Closure to apply.

Without Closure, the contingent space remains open. A fully characterized contingent space assigns no weights, selects no realizations, and generates no observational predictions. Magnitude provides the structure within which Closure operates; it does not provide the operation itself. A bridge theory complete at Partition and Magnitude but lacking Closure knows the shape of the space but cannot say what happens in it.

**2.6 Reduction**

On this account, reduction means that the inter-level map is approximately injective in the ways that matter for explanation. A higher-level construct is reduced when different observational values reliably pick out distinct regions of the underlying state space, and when those regions are sufficiently well-structured that the required closure rules can be justified within the bridge theory. Reduction comes in degrees: it depends on how consistently distinct observational descriptions correspond to distinct dynamical regions.

Temperature—the successful case—picks out distinct regions of phase space, well characterized by the Liouville measure, and the required closure rules, the microcanonical ensemble, are well-motivated within the bridge theory. Caloric, the failed case, collapsed at Partition: its equivalence classes were not stable under Hamiltonian dynamics, and the bridge theory failed before reaching Magnitude or Closure (Chang 2004).

Reduction is therefore a property diagnosed across all three completeness levels. At Partition, it requires that the chosen coarse-graining produces equivalence classes that are stable under the dynamical flow—that distinct observational descriptions reliably track distinct dynamical regions. At Magnitude, it requires that those regions be well-characterized by the structural measure, so that the contingent space has a determinate shape. At Closure, it requires that the closure rules needed to assign weights or select realizations are derivable within the bridge theory rather than imported from outside it. Failure under any of the conditions blocks reduction, but the level of failure matters: it determines the kind of theoretical work needed to achieve reduction.

## 3. The Mirror Test

### 3.1 Two Types of Closure Rule

Pearl's (2000) "do" operator captures an operation structurally similar to a closure rule. A closure rule is a surgical intervention on the contingent space: it modifies the set of admissible dynamical states without altering the dynamics themselves. The Stosszahlansatz, for example, does not move molecules—it specifies which dynamical states are admitted as valid realizations of the bridge-level description by imposing pre-collision statistical independence. The time-asymmetry is introduced into the representation, not into the dynamics. This filtering is precisely why Loschmidt-reversed trajectories remain fully admissible at the dynamical level while being excluded from the bridge-level description.

Two dynamically equivalent closure rules can nonetheless differ in a structurally fundamental way—one that turns on whether the elements they select are invariant under the symmetry group of the dynamical theory—the group of automorphisms that preserve its equations of motion. In Hamiltonian mechanics, this group includes time reversal; in Schrödinger evolution, it includes unitary transformations. A closing rule selects a subset or distribution invariant under this group; an introducing rule selects one that breaks that invariance. This distinction determines whether the emergence generated by a closure rule is provisional or permanent relative to the dynamical theory.

### 3.2 The Mirror Test

The distinction between closing and introducing rules is easy to state but hard to apply without a criterion. The same closure rule can be described in ways that make it look like either. The Stosszahlansatz imposes pre-collision statistical independence—a condition that, stated abstractly, could be a structural constraint on the collision representation or a selection rule derivable from the dynamics. Intuition about which programs are "really" making symmetry-breaking assumptions is unreliable: the history of the arrow of time debate is a record of confident programs that turned out to be attempting the impossible. The criterion has to operate on the structure of the closure rule itself, independently of how it is described and independently of what derivations have been attempted.

The Mirror Test formalizes the distinction between closing rules and introducing rules. Let G be the symmetry group of the dynamical theory—the automorphisms that preserve its equations of motion, fixed independently of any closure rule. Let $C(x)$ be the contingent space for observational description x, and let $S(\Sigma)$ be the subset or distribution over $C(x)$ selected by closure rule $\Sigma$. The Mirror Test asks: is $S(\Sigma)$ invariant under G?

A closure rule passes the Mirror Test if $S(\Sigma)$ is invariant under G: for every g in G, $g \cdot S(\Sigma) = S(\Sigma)$. It is then a closing rule, and the emergence it generates is provisional relative to G, as the symmetry group provides no structural barrier to closure.

A closure rule fails the Mirror Test if there exists at least one g in G such that $g \cdot S(\Sigma) \neq S(\Sigma)$. In that case, it is an introducing rule, and the emergence it generates is permanent relative to G, as no development of the dynamical theory that preserves G can supply a selection that breaks it.

What the test can establish depends on both features. G is fixed by the automorphism structure of the dynamical theory, independently of any closure rule and independently of what derivations have been attempted: the verdict is structural, not a report on the current state of the literature. And permanence is always stated relative to G—the Mirror Test identifies permanent emergence within a specific dynamical class, not emergence in every possible theoretical context. These are not caveats. They are what make the criterion objective and its verdicts precise.

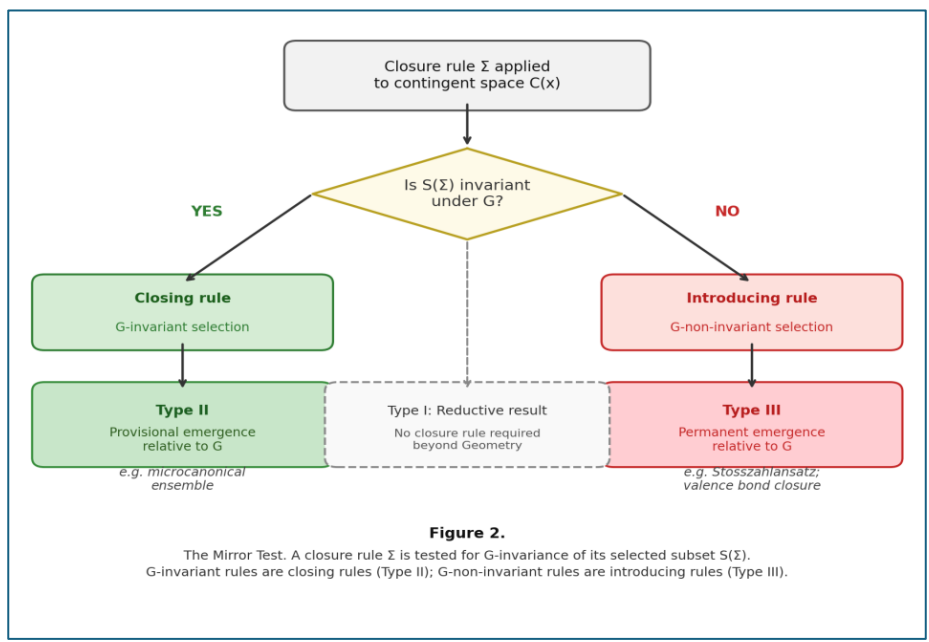

Figure 2.
The Mirror Test. A closure rule $\Sigma$ is tested for G-invariance of its selected subset $S(\Sigma)$. G-invariant rules are closing rules (Type II); G-non-invariant rules are introducing rules (Type III).

### 3.3 Paradigm Cases

The microcanonical ensemble assigns uniform weight over the energy surface—a distribution invariant under time reversal: reversing all molecular momenta leaves the energy surface and its Liouville measure unchanged. The closure rule passes the Mirror Test, and the emergence it generates is provisional relative to G.

The Stosszahlansatz, an introducing rule, imposes pre-collision statistical independence. Under time reversal, pre-collision states map to post-collision states and vice versa, so the subset of trajectories satisfying pre-collision independence is not preserved: its time-reverse selects trajectories satisfying post-collision independence, a different and incompatible subset. The Loschmidt (1876) theorem independently corroborates this verdict: the time-asymmetric selection is provably absent from any dynamics belonging to the Hamiltonian symmetry class. Albert (2000) argues that recovering thermodynamic asymmetry requires a Past Hypothesis constraining initial conditions—a constraint that, on the present framework, is precisely an introducing rule operating at Closure.

The valence bond closure rule selects localized electron pairs between specific atomic centers—and the relevant symmetry group is electron permutation symmetry, a symmetry of the Hamiltonian for any system of identical particles (Bader, 1990; Hoffmann et al., 2003). The point is not that the underlying quantum state lacks permutation symmetry, but that the bond-constituting closure rule selects electron configurations that are not invariant under permutation of identical particles. Permuting two electrons assigned to a bond region with electrons on an adjacent atom maps the selected subset to a different subset, one the closure rule did not select. The valence bond closure rule fails the Mirror Test relative to G = electron permutation symmetry. The directed, localized character of chemical bonds is permanently emergent relative to the unitary symmetry of quantum mechanics—and the Mirror Test verdict here has nothing to do with time.

Phase transitions are the diagnostically important borderline case. Under the ensemble interpretation, the closure rule assigns the maximum-entropy distribution over microstates compatible with the macrostate constraints—a distribution invariant under time reversal, and so a closing rule: the emergence of critical behavior is provisional. Under the spontaneous symmetry breaking interpretation, the closure rule selects a specific pure state from among the degenerate ground states at the critical point, breaking the rotational or translational symmetry of the Hamiltonian—an introducing rule whose emergence is permanent relative to G. The Mirror Test does not adjudicate between these interpretations—that is domain-specific work. What it identifies is precisely what is at stake: whether the novelty is provisional or permanent.

## 4. A Tripartite Taxonomy of Emergence

The weak/strong binary that dominates the emergence literature has a specific failure: it cannot tell you whether further work on a program could succeed. A phenomenon classified as weakly emergent is, on that account, derivable in principle—but the classification carries no information about whether any particular derivation strategy is structurally capable of achieving it. A century of derivation programs aimed at the arrow of time were, on the weak/strong account, making progress on a merely weak emergence problem. The Mirror Test shows they were attempting the impossible. That is not a failure of ingenuity. It is a structural verdict. What is needed is a taxonomy that carries that information—one whose categories track not just what has been derived but what can be derived, given the symmetry structure of the dynamical theory.

The Mirror Test yields exactly three cases.

*Type I — Reductive Result.*

The inter-level map is approximately injective, and the closure rules are derivable within the bridge theory. Nothing in the symmetry group G blocks the derivation. Temperature is the paradigm: the microcanonical ensemble passes the Mirror Test, and the reduction is essentially complete.

*Type II — Provisional Emergence.*

The bridge theory requires a closure rule that the dynamical theory itself does not provide, but the chosen distribution still passes the Mirror Test. The observational property is novel, yet G imposes no structural obstacle to eventual derivation. Phase transitions under the ensemble interpretation fit this pattern: the Mirror Test says YES, and the emergence is provisional, pending a finite-system account.

*Type III — Permanent Emergence relative to G.*

The bridge theory requires an introducing rule whose selected subset fails the Mirror Test. The required structure is absent from the dynamical theory, as constrained by G. Thermodynamic irreversibility via the Stosszahlansatz is the temporal paradigm; the directed chemical bond in valence-bond theory is the non-temporal case; and phase transitions under the spontaneous-symmetry-breaking interpretation form the borderline case, where the verdict is NO.

The three types are not descriptive categories applied after the fact. They are determined by a single formal question—does the required closure rule pass the Mirror Test?—applied to independently fixed features of the bridge theory.

## 5. Underdetermination and Constrained Pluralism

### 5.1 The Underdetermination Condition

The dynamical theory fixes the state space over which the inter-level map is defined—but not the map itself. When multiple Partition choices each generate a contingent space that can be completed at all three levels, and none is privileged by the dynamical theory, the bridge theory is structurally underdetermined. No further development of the dynamical theory will resolve it, because the inter-level map belongs to the bridge theory, not to the dynamical theory.

Chemical bonding makes this concrete. Quantum mechanics determines the electron density but does not specify how it is partitioned into bond contributions. Molecular orbital theory, valence bond theory, atoms-in-molecules, and natural bond orbital analysis each provide a different partition, generating distinct contingent spaces for the same quantum state. All four are internally complete at all three levels. None is privileged by quantum mechanics (Chang 2012; Bader 1990; Hoffmann et al. 2003; Hendry 2010).

## 5.2 Constrained Pluralism

When structural underdetermination obtains, the appropriate response is constrained pluralism: multiple bridge theories, each complete at all three levels, coexist and are deployed where their Partition fits the specific explanatory demand. Pluralism is permissible. It is not unlimited.

What makes the pluralism permissible is that the dynamical theory genuinely underdetermines the Partition. What constrains it is that bridge theories failing completeness at any level are not adequate, regardless of their predictive success. Adequacy is task-relative without being arbitrary: different Partition choices serve different explanatory demands, and the completeness conditions that make a framework adequate are fixed.

Chang's (2012) defense of scientific pluralism arrives at a similar conclusion, and the proximity is instructive. Chang holds that multiple frameworks are legitimate as long as each is useful—a pragmatist criterion that licenses pluralism without constraining it. But "useful" cannot distinguish between a bridge theory that is adequate and one that merely predicts well while failing a completeness condition it has not noticed. Molecular orbital theory and valence bond theory are both useful by any pragmatist measure; what distinguishes them is not utility but the structural question of which Partition each fixes and whether the closure rules each employs are completing a determinate contingent space or papering over an unresolved one. The present account replaces the pragmatist criterion with a structural one. Pluralism is warranted when the dynamical theory architecturally underdetermines the Partition and each competing bridge theory satisfies all three completeness conditions. Where those conditions are not met, the pluralism is not principled but merely convenient. Kellert, Longino, and Waters (2006) and Dupré (1993) develop the broader case for pluralism; the bridge theory framework provides the structural account of what makes it warranted and what constrains it.

# 6. Illustrations

## 6.1 Statistical Mechanics

Statistical mechanics connects Hamiltonian mechanics to thermodynamics via an inter-level map $\pi$ that sends molecular configurations—complete specifications of the positions and momenta of approximately $10^{23}$ particles—to macrostates characterized by temperature, pressure, and volume. The contingent space $C(M) = \Gamma_M$ is the region of phase space compatible with macrostate M: astronomically large, and the dominant structural feature of the bridge theory, not a correction to it.

The coarse-graining into macrostates defined by temperature, pressure, and volume is substantially stabilized, satisfying projectability, differentiability, and closure under iteration (Boltzmann 1877). The Liouville measure on $\Gamma_M$ is singled out by the Hamiltonian dynamics; the contingency measure $\log(|\Gamma_M| / v_0)$ is defined here, before any closure rule is applied. The reference cell $v_0$ must be supplied as a bridge-theoretic construct; contingency measure differences, $v_0$-independent by construction, suffice for all comparative thermodynamic claims.

The bridge theory requires two structurally distinct closure rules whose difference the Mirror Test makes precise. The microcanonical ensemble assigns uniform weight to the energy surface; its distribution is invariant under time reversal, so it is a closing rule generating provisional emergence. The Stosszahlansatz selects a subset of trajectories satisfying pre-collision independence; that subset is not preserved under time reversal, so it is an introducing rule generating permanent emergence relative to the Hamiltonian dynamical class. Any account claiming that thermodynamic behavior follows from Hamiltonian mechanics while leaving either closure rule unexamined has not completed the bridge theory.

The typicality program occupies a third structural position, distinct from both. Typicality arguments establish that thermodynamic behavior holds for almost all microstates in $\Gamma_M$—conclusions formulated as ratios of contingent space volumes in which $v_0$ cancels. The typicality program therefore does not require a fixed $v_0$, and its conclusions are $v_0$-independent by construction. It is not a rival to the treatment of $v_0$ as a genuine bridge-theoretic construct: typicality addresses what holds for the overwhelming majority of the contingent space; the $v_0$ question concerns the scale of that space. The two programs operate at different levels of the bridge theory. Similarly, the typicality program and the Stosszahlansatz are not rivals: typicality establishes what holds for almost all microstates under the Liouville measure, while the Stosszahlansatz selects a specific subset of trajectories by imposing pre-collision independence. They occupy different structural positions within Closure, not competing answers to the same question.

**6.2 Quantum Chemistry**

Chemical bonds are not discovered in the quantum state. The Partition choice constitutes them. This is what the persistent underdetermination of chemical bonding reveals, and what the framework explains.

Quantum chemistry connects quantum mechanics to structural chemistry via an inter-level map $\pi$ that sends quantum states to structural chemical descriptions—bond types, molecular geometry, reactivity. Quantum mechanics fixes the electron density but does not say how that density constitutes a bond. Molecular orbital theory, valence bond theory, atoms-in-molecules, and natural bond orbital analysis each define a different $\pi$ and generate a distinct contingent space for the same quantum state. All four are adequate within their respective domains; none is privileged by quantum mechanics (Bader 1990; Hoffmann et al. 2003; Chang 2012; Hendry 2010). The Hilbert space measure is well-defined once Partition is specified—what is contested is not the geometry of the contingent space but which contingent space the geometry is of.

The closure rules each analysis scheme employs are introducing rules: the directed, localized bond properties they constitute break the permutation symmetry of quantum mechanics and fail the Mirror Test relative to G. This is why the bonding debate cannot be resolved by better quantum mechanics. The underdetermination is at Partition, and the dynamics do not fix the Partition. The appropriate response is constrained pluralism: molecular orbital theory where delocalized electron behavior is explanatorily relevant; valence bond theory where bond directionality matters; atoms-in-molecules where atomic contributions are the target. Each is structurally complete within its domain. None is globally privileged.

### 6.3 Molecular Genetics

Seventy years of molecular knowledge have not produced a stable definition of the gene. The framework explains why: the dispute is not about the molecular biology, which is well-characterized, but about which features of the molecular biology count as "the gene" for a given explanatory purpose—a Partition question that the dynamics do not answer.

Molecular genetics connects molecular biology to Mendelian genetics via an inter-level map $\pi$ that maps states of the molecular substrate—DNA sequences, transcription factors, regulatory elements, and RNA molecules—to Mendelian genetic descriptions: genes, alleles, traits, and phenotypes. Multiple coarse-grainings compete: open reading frames, transcription units, regulatory regions, and genomic regions with heritable phenotypic effects. Each satisfies the completeness conditions for its respective explanatory demands; none is globally privileged by the underlying molecular biology (Waters 1994; Griffiths and Neumann-Held 1999). The molecular substrate is well-characterized, and a Magnitude is available once Partition is fixed. The dispute is not about the geometry of the contingent space. It is about which contingent space to define.

The Mirror Test cannot yet be applied to Closure here, because the test requires a settled Partition to evaluate: without a fixed $\pi$, there is no determinate contingent space, and without a determinate contingent space, there is no fixed subset for the test to evaluate. This is not a failure of the framework—it is a precise diagnosis of where the work remains. Further molecular knowledge will not resolve the debate over the gene concept. The dynamics do not determine the Partition, and different choices will remain adequate for different explanatory demands.

## 7. Conclusion

That the inputs to the bridge theory resist derivation has a structural source that existing accounts of reduction and emergence have not located: the architectural position of the bridge theory itself. A third theoretical role, distinct from both the dynamical and observational theories it connects, generates an object neither theory has a slot for—the contingent space—and requires three completeness conditions whose ordering is a constraint imposed by the architecture, not a convention.

The contingency measure—$\log[(1/v_0) \int \delta(\pi(\gamma) - x) \, d\gamma]$—makes the two-component structure of Magnitude explicit. The Liouville measure $d\gamma$ is fixed by the dynamical theory's own dynamics and is not a bridge-theoretic choice. The reference cell $v_0$ is a genuine bridge-theoretic construct that the dynamical theory does not supply. Differences in contingency measures are $v_0$-independent and always well-defined; absolute values are conventional unless a more fundamental framework resolves $v_0$. The contingency measure makes explicit what Boltzmann's formula already measured: the pre-probabilistic volume of the contingent space. The two-component structure—$d\gamma$ from the dynamical theory, $v_0$ as a genuine bridge-theoretic construct—is implicit in Boltzmann and explicit here, extended now to any inter-level map, not just to the phase-space partition of statistical mechanics.

The Mirror Test introduces a formal criterion for distinguishing two structurally distinct types of closure rule. Closing rules select a G-invariant subset or distribution; their emergence is provisional relative to G. Introducing rules select a G-non-invariant subset or distribution; their emergence is permanent relative to G. The Stosszahlansatz is the paradigm introducing rule in the temporal case; the valence bond closure rule is the non-temporal case. Phase transitions discriminate between provisional and permanent emergence depending on which closure rule interpretation is correct—exactly the kind of tractable dispute the Mirror Test is designed to locate.

The tripartite taxonomy follows from the formal criterion. Type I reductive results arise when the inter-level map is approximately injective and closure rules are derivable within the bridge theory. Type II provisional emergence arises when closing rules are required, but their selected distributions are G-invariant. Type III permanent emergence arises when introducing rules are required. Constrained pluralism is the correct structural response to Partition underdetermination—licensed only when the dynamical theory architecturally underdetermines the inter-level map, and each competing bridge theory satisfies all three completeness conditions.

Three disputes that have resisted resolution for decades are diagnosed precisely by the framework. The debate over the gene concept is a problem of Partition stabilization: further molecular knowledge will not resolve it because the dynamics do not determine the Partition. The chemical bonding debate is a dispute over Partition underdetermination, with permanent introducing rules at Closure: the four analysis schemes are not competitors for the same truth but structurally complete answers to different questions. The foundations of statistical mechanics require both a closing rule generating provisional emergence and an introducing rule whose permanence is established by an independent theorem: the typicality program and the Stosszahlansatz are not rivals, but occupants of different structural positions, and neither displaces the other's role in the bridge theory. In each case, the framework does not resolve the dispute—that is, domain-specific work. What it provides is something prior: a precise diagnosis of where the dispute lies and what any resolution must provide.